\documentstyle[12pt]{article}
\title{\huge  Wetting of an Ising system with perfect and corrugated surfaces in a transverse field} 
\author{\bf  
 L. Bahmad, A. Benyoussef, and H. Ez-Zahraouy 
\\
 Laboratoire de Magn\'{e}tisme et de la Physique
 des Hautes Energies
\\
Universit\'{e} Mohammed V, Facult\'{e} des Sciences \\
 Avenue Ibn Batouta, B.P. 1014, Rabat, Morocco
}
\date{ }
\begin{document}
\maketitle 
\date{}
\begin{abstract}
\mbox{~~~  } Using the mean field theory, a comparative study of the wetting and layering transitions of a spin-$1/2$ Ising model with perfect and corrugated surfaces, is established. The phase diagrams are investigated and compared in the presence of both a longitudinal and surface fields. The effect of increasing the temperature and the transverse field on the wetting and layering transitions is outlined.
 
\end{abstract}
 
\noindent ----------------------------------- \newline
\newpage
\section{Introduction}

\mbox{  } Recently, the wetting and layering transitions of magnetic Ising systems have been subject of great interest. Experimental studies have motivated much theoretical works in order to understand and explain the growth of thin layers from only single atoms. A simple lattice-gas model with layering transitions has been introduced and studied in the mean field approximation by de Oliveira and Griffiths [1].
Multilayer films adsorbed on attractive substrates may exhibit a variety of possible phase transitions, as has been reviewed by Pandit {\it et al.} [2], Nightingale {\it et al.} [3], Patrykiejew {\it et al.} [4] and Ebner {\it et al.} [5-8]. One type of transitions is the layering transitions, in which the thickness of a solid film increases discontinuously by one layer as the pressure is increased. Such transitions have been observed in a variety of systems including for example $^{4}He$ [9,10] and ethylene [11,12] adsorbed on graphite. Ebner and Saam [13] carried out Monte Carlo simulations of such a lattice gas model. Huse [14] applied renormalization group technique to this model. It allowed the study of the effects on an atomic scale in the adsorbed layers. The lattice gas models applied to the wetting phenomena was reviewed by Dietrich [15]. The effect of finite size on such transitions has been studied, in a thin film confined between parallel planes or walls, by Nakanishi and Fisher [16] using mean field theory. \\
The model of transverse field was originally introduced by de Gennes [17] for hydrogen-bonded ferroelectrics such as $KH_{2}PO_{4}$. Since then, this model has been applied to several physical
systems, like $DyVO_{2}$, and studied by a variety of sophisticated
techniques [18-21]. 
In order to study the amorphization of a crystalline Ising ferromagnet with a transverse field, Kaneyoshi [22] introduced the differential operator technique as a generalised but approximate Callen relation derived by Sa Barreto and Fittipaldi [23]. The system has a finite transition temperature, which can be decreased by increasing the transverse field to a critical value $\Omega_{c}$. The effect of a transverse field on the critical behaviour and the magnetisation curves was studied [18-21] and by Kaneyoshi {\it et al.} [24,25]. Using the perturbative theory, Harris {\it et al.} [26] have studied the layering transitions at $T=0$ in the presence of a transverse field.
Benyoussef and Ez-Zahraouy have studied the layering transitions of Ising model thin films using a real space renormalization group [27], and transfer matrix [28] methods. In a previous work [29] we have established the effect of temperature and transverse field, for a fixed surface field values, on the wetting and layering transitions of an Ising system with perfect surfaces.   
The purpose of this work is to study the effect of a corrugated surface on the wetting and layering transitions, of an Ising system spin-$1/2$, using the mean field theory. The outline of this work is as follows. In Section $2$, the model and method are presented. Section $3$ is devoted to study the ground state for the perfect and corrugated surface cases. The results and the phase diagrams obtained are discussed in Section $4$.
 
\section{Model and method}

\mbox{~~~} We consider a film with finite thickness of a three dimensional spin-$1/2$ model, with $N$ layers, limited by two surfaces. Figure 1 represents the geometry of this system for \\
 $a)$ two perfect surfaces, \\
 $b)$ presence of a corrugated surface. \\
The Hamiltonian describing this system can be written as
\begin{equation}
{\cal H}=-\sum_{<i,j>}J_{ij}S^{z}_{i}S^{z}_{j}-\sum_{i}(H_{i}S^{z}_{i}+\Omega S^{x}_{i})
\end{equation}
where the first summation is carried out over nearest-neighbour pairs of spins. $S^{\alpha}_{i}=\pm 1$, $(\alpha=z,x)$ are the Pauli matrices for a spin $i$. The exchange interaction $J_{ij}=J$ is assumed to be constant, and $\Omega$ is the transverse field.
The total longitudinal field $H_{i}$, applied on a site $i$, is distributed  according to the nature of the top surface as it is discussed below. \\
For a system with perfect surfaces, Fig. 1a, $H_{i}$ is assumed to be uniform in a layer $k$, and defined by: 
\begin{equation}
H_{k}=\left\{ 
\begin{array}{lll}
H+H_{s1} & \mbox{for} & k=1 \\ 
H & \mbox{for} &  1 < k < N  \\ 
H+H_{s_{N}} & \mbox{for} & k=N
\end{array}
\right.
\end{equation}
where the surface fields $H_{s1}$ and $H_{s_{N}}=-H_{s1}$ are applied on the first layer $k=1$ and the last layer $k=N$, respectively. $H$ is the external field applied on the whole system. \\
For a corrugated surface film with n steps, each step containing L spins in the $x$ direction, $H_{i}$ is given by: 
\begin{equation}
H_{i}=\left\{ 
\begin{array}{lll}
H+H_{s1} & \mbox{for} & i \epsilon S_{1} \\ 
H & \mbox{for} &  i \epsilon S_{k} \mbox{:  } 1<k<N  \\ 
H+H_{s_{N}} & \mbox{for} & i \epsilon S_{N}
\end{array}
\right.
\end{equation}
where the surface field $H_{s1}$ acts on the corrugated surface $S_{1}$, and $H_{s_{N}}=-H_{s1}$ is applied on the bottom surface $S_{N}$, Fig. 1b. The configurations $S_{k} (0\leq k \leq N)$ are defined in Section 3 for both the perfect and corrugated surfaces. $H$ is the external field applied on the whole system. \\
The part of the Hamiltonian describing a site $i$, for the perfect and corrugated surfaces, is given by 
\begin{equation}
{\cal H}_{i}=(\sum_{j\ne i}J_{ij}S^{z}_{i}+H_{i})S^{z}_{i}+\Omega S^{x}_{i}
\end{equation}
the summation runs over nearest neighbour sites $j$ of $i$. The diagonalization of the operator ${\cal H}_{i}$ leads to the eigen values $\lambda_{i}^{\pm}=\pm\sqrt{x_{i}^{2}+\Omega^{2}}$, where $x_{i}=\sum_{j\ne i}J_{ij}S^{z}_{i}+H_{i}$. \\   
The magnetisation per site can be written as 
\begin{equation}
  m_{i}=\frac{TrS_{i}^{z}\exp(-\beta H_{i})}{Tr\exp(-\beta H_{i})}=[\frac{(k_{i}+\lambda_{i}^{+})^{2}-\Omega^{2}}{(k_{i}+\lambda_{i}^{+})^{2}+\Omega^{2}}]\tanh(\beta \sqrt{x_{i}^{2}+\Omega^{2}}).   
\end{equation} 
Using the mean field theory, the free energy of the system can be written as follows  
\begin{equation}
      F=-\frac{1}{\beta}\sum_{i}\log(2\cosh(\beta\lambda_{i}^{+}))+
\frac{1}{2}\sum_{i}\sum_{j}J_{ij}m_{i}m_{j} .
\end{equation}
For a simple cubic lattice, the magnetisation and the free energy for each plane $k$ are given, respectively, by the following expressions:
\begin{equation}
m_{k}=\frac{1}{\lambda_{k}}[4m_{k}+(m_{k+1}+m_{k-1}+H^{'})]\tanh(\beta^{'}\lambda_{k})  
\end{equation}
and
\begin{equation}
F_{k}=-\frac{1}{\beta}\log(2\cosh(\beta  \lambda_{k}))+\frac{1}{2}m_{k}(4m_{k}+(m_{k+1}+m_{k-1})  
\end{equation}
with 
\begin{equation}
 \lambda_{k}=\sqrt{ (4m_{k}+(m_{k+1}+m_{k-1})+H^{'})^2+\Omega^{'2}},  
\end{equation}
assuming that $J_{ij}=J$, the reduced parameters are defined by:  
$\beta^{'}=\beta J$, $H^{'}=\frac{H}{J}$, $\Omega^{'}=\frac{\Omega}{J}$. 
 
\section{Ground state}
\subsection{Perfect surface}
\mbox{~~~} The ground state of this model is established in Fig. 2a. The start point is a situation where all spins are down: configuration  $S_{0}$.
For $H_{s1}/J\leq 1$, we have only one transition, $S_{0}\leftrightarrow S_{N}$. By contrast, for $H_{s1}/J\geq 1$, we have three transitions namely; the first transition, $S_{0}\leftrightarrow S_{1}$, arises for $H/J=1-H_{s1}/J$ , the second transition, $S_{1}\leftrightarrow S_{N-1}$, occurs for $H/J=0$, while the last transition $S_{N-1}\leftrightarrow S_{N}$, is seen for $H/J=1+H_{s1}/J$. The notation $S_{k}(k=0,1,...,N)$ is a situation where, starting from the top surface, k layers are spin-up and the remainder $N-k$ layers are spin-down.
The Fig. 2a shows that the transitions arise for values of $H_{s1}/J$ above a critical value, $H^{c}_{s1}/J=1$. 

\subsection{Corrugated surface}
\mbox{~~~} The model is described in section 2, the corresponding ground state is illustrated in Fig. 2b for a system formed with $N=20$ layers, $n=2$ steps and $L=10$ spins, for each step, in the $x$ direction. The start point is a situation, called $S_{0}$, where all the spins are down. Under the effect of increasing the surface field $H_{s1}/J$, and for a bulk field value $H/J$ close to $1-H_{s1}/J$, the spins of the top external surface, called $S_{1}$, will flip and become up. Increasing the bulk field, the spins  of the layer 2 will flip, this situation is noted $S_{2}$. Increasing the bulk field more and more, the spins of the layer 3 will flip, this configuration is noted $S_{3}$, and so on. As it is shown Fig. 2b, the transition $S_{0} \leftrightarrow S_{1}$, which appears for $H/J=1-H_{s1}/J$ and the transition $S_{N-1} \leftrightarrow S_{N}$, arising for $H/J=-1+H_{s1}/J$, are fixed only by the surface field $H_{s1}/J$ value, and do not depend on the parameters $n$ and $L$. By contrast, the number of transitions parallel to the line $H/J=0$ is close to the number of corrugated surfaces $n$, and the corresponding value of $H/J$ is fixed by the number of spins per step $L$.
A generalisation of the ground state energy $E(S_{k},n,L,N)$, of this system, is given in Appendix A, as a function of arbitrary values of the parameters: $n,L$ and $N$. 

\section{Results and discussion}
\mbox{  }Numerical results are giving for a system formed with $N=20$ ferromagnetic layers of a spin$-1/2$ Ising model with free bound conditions. In order to examine the temperature effect, in absence of transverse magnetic field, $\Omega /J=0$, on the wetting and layering transitions, we plot in Fig. 3 the corresponding phase diagrams. By increasing the temperature we show the existence of a critical value, $T_{w} /J$, above which a sequence of layering transitions occurs this means that the temperature effect is sufficient to produce layering transitions. This temperature, $T_{w} /J$, is called the "wetting temperature", which depends on the transverse field and surface magnetic field values. The first and last layer transitions occur in the absence of the transverse magnetic field, but only under the increasing bulk magnetic field for a temperature $T/J > T_{w} /J$, the other layering transitions arise for increasing temperatures. On the other hand, the corrugation effect is to accelerate the layering transitions phenomenon since the transitions $S_{0} \leftrightarrow S_{2}$ and $S_{0} \leftrightarrow S_{3}$ are present in the corrugated surface case: Fig. 3b and are absent in the perfect surface situation: Fig 3a. The same behaviour is found when inverting the temperature and the transverse field roles, as it is summarised in Fig. 4. The phase diagrams show that the layer transitions are found for $\Omega /J > \Omega_{w} /J$ with increasing transverse magnetic field at fixed temperature. $\Omega_{w} /J$ is called the "wetting" transverse field.
The same transitions, found in Fig. 3, are still present for the corrugated surface case, Fig. 4b, and disappear for the perfect surface situation, Fig 4a. \\
Figs. 3 and 4 are plotted for a surface field value $H_{s1}/J=0.9$, less than the critical value $1$. The effect of increasing the surface field above the critical value is summarised in Figs. 5 and 6 for $H_{s1}/J=1.2$. It is worth to note that the "wetting" parameters, $T_{w} /J$ and $\Omega_{w} /J$, disappear since the layering transitions are present even for very low temperature: Figs 5a and 5b, and very low transverse field:  Figs 6a and 6b. Furthermore, the transitions $S_{1} \leftrightarrow S_{2}$ and $S_{2} \leftrightarrow S_{3}$ are present in Figs. 5b and 6b for $T/J \rightarrow 0$ and $\Omega/J \rightarrow 0$, respectively, in good agreement with the ground state illustrated in Fig. 2b, for the corrugated surface case. On the other hand, the dependency of the wetting temperature $T_{w} /J$ on the wetting transverse field $\Omega_{w} /J$, is illustrated in Fig. 7, corresponding to the surface field value $H_{s1}/J=0.9$, for the both cases: perfect and corrugated surfaces. As it has been outlined above, the corrugation effect is to accelerate the wetting phenomenon since the wetting region, of the phase diagram in Fig. 7, occurs first for the corrugated surface before that one  corresponding to the perfect surface.     
 
\section{Conclusion}

Within the mean field theory, we have studied the wetting and layer transitions of a 3-D spin$-1/2$ Ising model. For a surface field less than the critical value $1$, the system exhibits a sequence of layering transitions only above a wetting temperature (wetting transverse field) for fixed transverse field (temperature). For surface fields higher than the critical value, the layering transitions under only under the effect of increasing the bulk field even for $\Omega/J=0$ or $T/J \rightarrow 0$. Furthermore the corrugation effect is accelerate the layering transitions and wetting phenomena compared to perfect surfaces. \\
  
{\Large Acknowledgements:} \\
One of the authors L. B. would like to thank the support of the program: PARS Physique $035$.  
\newpage \noindent{\bf References}
\begin{enumerate}
\item[{[1]}] M. J. de Oliveira  and R. B. Griffiths , Surf. Sci. {\bf 71}, 687 (1978).
\item[{[2]}] R. Pandit, M. Schick and M. Wortis, Phys. Rev. B {\bf 26}, 8115 (1982).
\item[{[3]}] M. P. Nightingale, W. F. Saam and M. Schick, Phys. Rev. B {\bf 30},3830 (1984).
\item[{[4]}] A. Patrykiejew A., D. P. Landau and K. Binder, Surf. Sci. {\bf 238}, 317 (1990).
\item[{[5]}] C. Ebner, C. Rottman and M. Wortis, Phys. Rev. B {\bf 28},4186  (1983).  
\item[{[6]}] C. Ebner and W. F. Saam, Phys. Rev. Lett. {\bf 58},587 (1987).
\item[{[7]}] C. Ebner and W. F. Saam, Phys. Rev. B {\bf 35},1822 (1987).
\item[{[8]}] C. Ebner, W. F. Saam and A. K. Sen, Phys. Rev. B {\bf 32},1558 (1987).
\item[{[9]}] S. Ramesh and J. D. Maynard, Phys. Rev. Lett. {\bf 49},47 (1982).
\item[{[10]}] S. Ramesh, Q. Zhang, G. Torso and J. D. Maynard, Phys. Rev. Lett. {\bf 52},2375 (1984).
\item[{[11]}] M. Sutton, S. G. J. Mochrie  and R. J. Birgeneou, Phys. Rev. Lett. {\bf 51},407 (1983);\\
S. G. J. Mochrie, M. Sutton, R. J. Birgeneou, D. E. Moncton and P. M. Horn, Phys. Rev. B {\bf 30},263 (1984).
\item[{[12]}] S. K. Stija, L. Passel, J. Eckart, W. Ellenson and H. Patterson, Phys. Rev. Lett. {\bf 51},411 (1983).
\item[{[13]}] C. Ebner and W. F. Saam, Phys. Rev. A {\bf 22},2776 (1980);\\
     ibid, Phys. Rev. A {\bf 23},1925 (1981);\\
     ibid, Phys. Rev. B {\bf 28},2890 (1983).
\item[{[14]}] D. A. Huse , Phys. Rev. B {\bf 30},1371 (1984). 
\item[{[15]}] S. Dietrich {\it Phase transitions and critical phenomena} Vol. 12, C. Domb and Lebowitz Eds., Academic press London and Orlando (1988).  
\item[{[16]}] H. Nakanishi and M. E. Fisher, J. Chem. Phys. {\bf 78},3279  (1983).
\item[{[17]}] P.G. de Gennes, Solid State Commun. {\bf 1}, 132 (1963).
\item[{[18]}]  A. Benyoussef, H. Ez-Zahraouy and M. Saber, Physica A, {\bf 198}, 593
(1993). 
\item[{[19]}]  A. Benyoussef and H. Ez-Zahraouy, Phys. Stat. Sol. (b) {\bf 179}, 521 (1993). 
\item[{[20]}]  A. Bassir, C.E. Bassir, A. Benyoussef, A. Klumper and J. Zittart, Physica A, {\bf 253}, 473 (1998).
\item[{[21]}]  A. Benyoussef, H. Ez-Zahraouy, J. Phys.: Cond. Matter {\bf 6}, 3411 (1994).    
\item[{[22]}]  T. Kaneyoshi, Phys. Rev. {\bf B 33}, 526 (1986).
\item[{[23]}]  F. C. Sa Barreto and I. P. Fittipaldi, Physica A, {\bf 129},
360 (1985).
\item[{[24]}]  T. Kaneyoshi, E. F. Sarmento and I. P. Fittipaldi, Phys. Stat.
Sol. (b) {\bf 150}, 261 (1988).
\item[{[25]}]  T. Kaneyoshi, E. F. Sarmento and I. P. Fittipaldi, Phys. Rev. 
{\bf B 38}, 2649 (1988).
\item[{[26]}]  A. B. Harris, C. Micheletti and J. Yeomans, J. Stat. Phys. 
{\bf 84},323 (1996).
\item[{[27]}]  A. Benyoussef and H. Ez-Zahraouy, Physica A, {\bf 206}, 196
(1994).
\item[{[28]}]  A. Benyoussef and H. Ez-Zahraouy, J. Phys. {\it I } France 
{\bf 4}, 393 (1994). 
\item[{[29]}]  L. Bahmad, A. Benyoussef, A. Boubekri, and H. Ez-Zahraouy,
Phys. Stat. Sol. (b) {\bf 215}, 1091 (1999). 
\end{enumerate}

\newpage \noindent{\bf Figure Captions}\newline
\newline
\noindent{\bf Figure 1.} Geometry of the model with $N$ layers in the $z$ direction. The system is infinite in the direction $y$ with: \\
a) perfect surfaces case, \\
b) presence of a corrugated surface with $n$ steps, each step contains $L$ spins in the $x$ direction.

\noindent{\bf Figure 2.} Ground state phase diagrams in the $(H/J,H_{s1} /J)$ plane for: \\ 
a) perfect surfaces case with $N=20$ layers, \\
b) corrugated surface case with $N=20$ layers, $n=2$ steps and $L=10$ spins for each step.

\noindent{\bf Figure 3} Phase diagrams in $(H/J,T/J)$ plane for $H_{s1}/J=0.9$ and $\Omega /J=0$  \\
a) perfect surfaces case, \\
b) presence of a corrugated surface.

\noindent{\bf Figure 4.} Phase diagrams in $(H/J,\Omega/J)$ plane for $H_{s1}/J=0.9$ and $T /J=0.5$  \\
a) perfect surfaces case, \\
b) presence of a corrugated surface.

\noindent {\bf Figure 5.} Phase diagrams in $(H/J,T/J)$ plane for $H_{s1}/J=1.2$ and $\Omega /J=0$  \\
a) perfect surfaces case, \\
b) presence of a corrugated surface.

\noindent {\bf Figure 6.} Phase diagrams in $(H/J,\Omega/J)$ plane for $H_{s1}/J=1.2$ and $T /J=0.5$  \\
a) perfect surfaces case, \\
b) presence of a corrugated surface.

\noindent {\bf Figure 7.} The wetting temperature $T_{w}/J$ dependency on the wetting transverse field $\Omega_{w} /J$, with fixed surface field $H_{s1}/J=0.9$, for the both cases: corrugated and perfect surfaces.
 
\newpage
\appendix 
{\huge Appendix } \\
{\Large Expression of the energy $E(S_{k},n,L,N)$: } \\
   In the following we will investigate, the ground state energy $E(S_{k},n,L,N)$ as a function of the parameters $S_{k},n,L$ and $N$ (see text). For $T=0$ the Hamiltonian of Eq. (1) allows to write :	 
\begin{equation}
E(S_{k},n,L,N)/J=A\times H_{s1}/J+B\times H/J+C. \\
\end{equation}
Where $A=N_{H_{s1}}$ is the coefficient of $H_{s1}$, $B=N_{H}$ is the coefficient of $H$ and $C=N_{J}$ is the coefficient of $J$, which depend on the parameters $S_{k},n,L$ and $N$ \\
For a given configuration $S_{k}, (x=0,1,...,N)$, of a corrugated surface system formed with $N$ layers, $n$ steps with $L$ spins for each step. These expressions will generate a sequence of transition lines from a configuration $S_{k}$ to a configuration $S_{l}$ when $0\leq k,l \leq N$. 
\subsection{Coefficients $N_{H_{s1}}(S_{k},n,L,N)$}
1- Configuration $S_{0}$ (all spins down):
 $N_{H_{s1}}(S_{0},n,L,N)=2n$.\\
2- Configuration $S_{k},(k=1, ..., N-1)$):
$N_{H_{s1}}(S_{k},n,L,N)=-2(n+1)(2L-1)$.\\ 
3- Configuration $S_{N}$ ( all spins up): 
$N_{H_{s1}}(S_{N},n,L,N)=-2n$.
\subsection{Coefficients $N_{H}(S_{k},n,L,N)$}
1- Configuration $S_{0}$:\\ $N_{H}(S_{0},n,L,N)=\sum_{z=1}^{n+1}(2z(L-1)+1)+(N-(n+1))(2(n+1)(L-1)+1)$.\\ 2- Configuration $S_{j}: 1 \leq j \leq n+1$:\\
$N_{H}(S_{j},n,L,N)=-\sum_{z=1}^{j}(2z(L-1)+1)+\sum_{z=j+1}^{n+1}(2z(L-1)+1-2L)+(N-(n+1))(2(n+1)(L-1)+1)$.\\ 
3- Configuration $S_{j}: n+1 < j \leq N$:\\ $N_{H}(S_{j},n,L,N)=-\sum_{z=1}^{n+1}(2z(L-1)+1)-(j-(n+1))(2(n+1)(L-1)+1)+(N-j)(2(n+1)(L-1)+1)$.
\subsection{Coefficients $N_{J}(S_{k},n,L,N)$}
1- Configurations $S_{0}$ :\\
$N_{J}(S_{0},n,L,N)=-(\sum_{z=1}^{n+1}(4(2z(L-1))+3)+(N-1-(n+1))(4(2(n+1)(L-1))+3)+(3(2(n+1)(L-1))+2))$.\\
2- Configuration $S_{1}$:\\
$N_{J}(S_{1},n,L,N)=-((4L+1)+\sum_{z=2}^{n}(2(4+2(L-2)+4(z(L-1)-L))+3)+(2(2(L-1)+4((n+1)(L-1)-L))+3)+\sum_{z=n+2}^{N-1}(4(2(n+1)(L-1))+3)+(3(2(n+1)(L-1))+2))$.\\
3- Configuration $S_{j}: 1 < j < n+1 $:\\
$N_{J}(S_{j},n,L,N)=-(\sum_{z=1}^{j-1}(5(2z(L-1)+1)-1)+\sum_{z=j}^{n}(2(4+2(L-2)+4(z(L-1)-L))+3)+(2(4+2(L-2)+4((n+1)(L-1)-L))+3)+(2(2(L-1)+4((n+1)(L-1)-L))+3)+\sum_{z=n+2}^{N-1}(4(2(n+1)(L-1))+3)+(3(2(n+1)(L-1))+2) )$. \\
4- Configuration $S_{n+1}$:\\
$N_{J}(S_{n+1},n,L,N)=-(\sum_{z=1}^{n}(5(2z(L-1)+1)-1)+(2(2(n+1)(L-1))+1)
+\sum_{z=n+2}^{N-1}(4(2(n+1)(L-1))+3)+(3(2(n+1)(L-1))+2) ) $. \\
5-Configuration $S_{j}: n+1 < j \leq N-1 $:\\
$N_{J}(S_{j},n,L,N)=-(\sum_{z=1}^{n}(5(2z(L-1)+1)-1)+\sum_{n+1}^{j-1}(4(2(n+1)(L-1))+3)+(2(2(n+1)(L-1))+1)+\sum_{z=j+1}^{N-1}(4(2(n+1)(L-1))+3)+(3(2(n+1)(L-1))+2) ).$\\
6-Configuration $S_{N}$:\\
$N_{J}(S_{N},n,L,N)=-(\sum_{z=1}^{n}(5(2z(L-1)+1)-1)+\sum_{n+1}^{N-1}(4(2(n+1)(L-1))+3)+(3(2(n+1)(L-1))+2) ) $
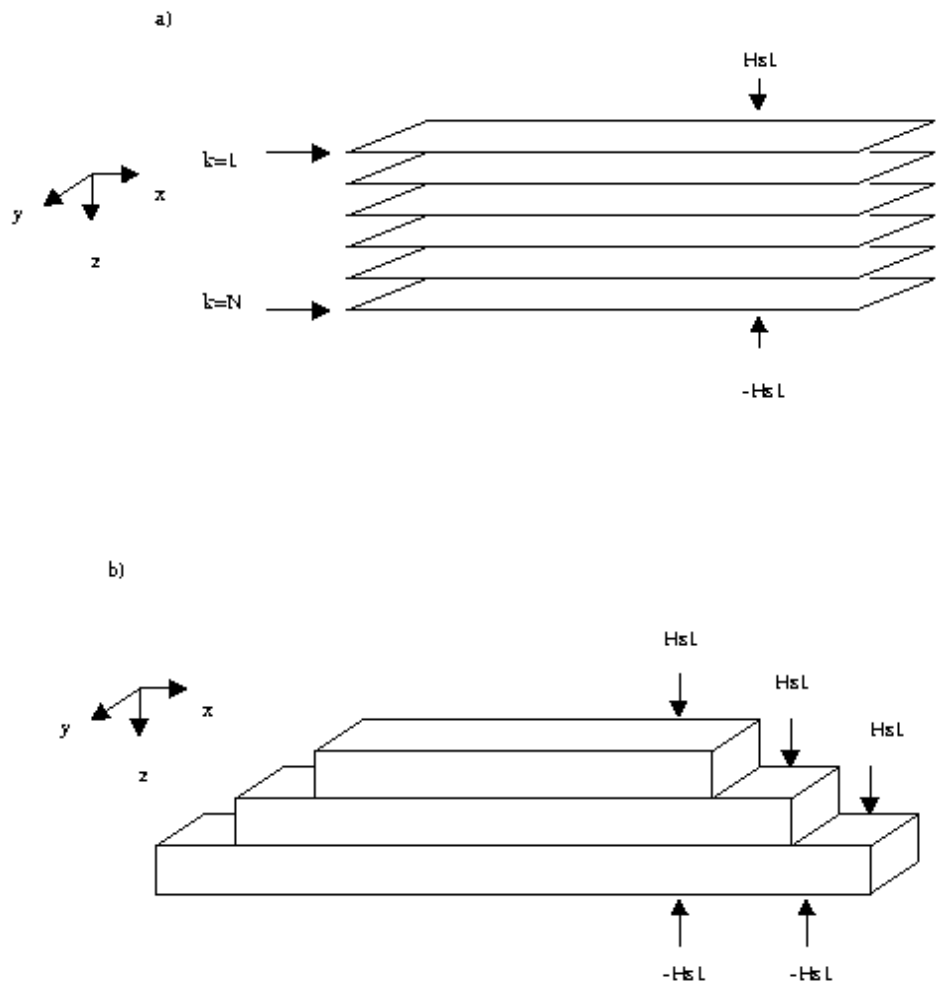
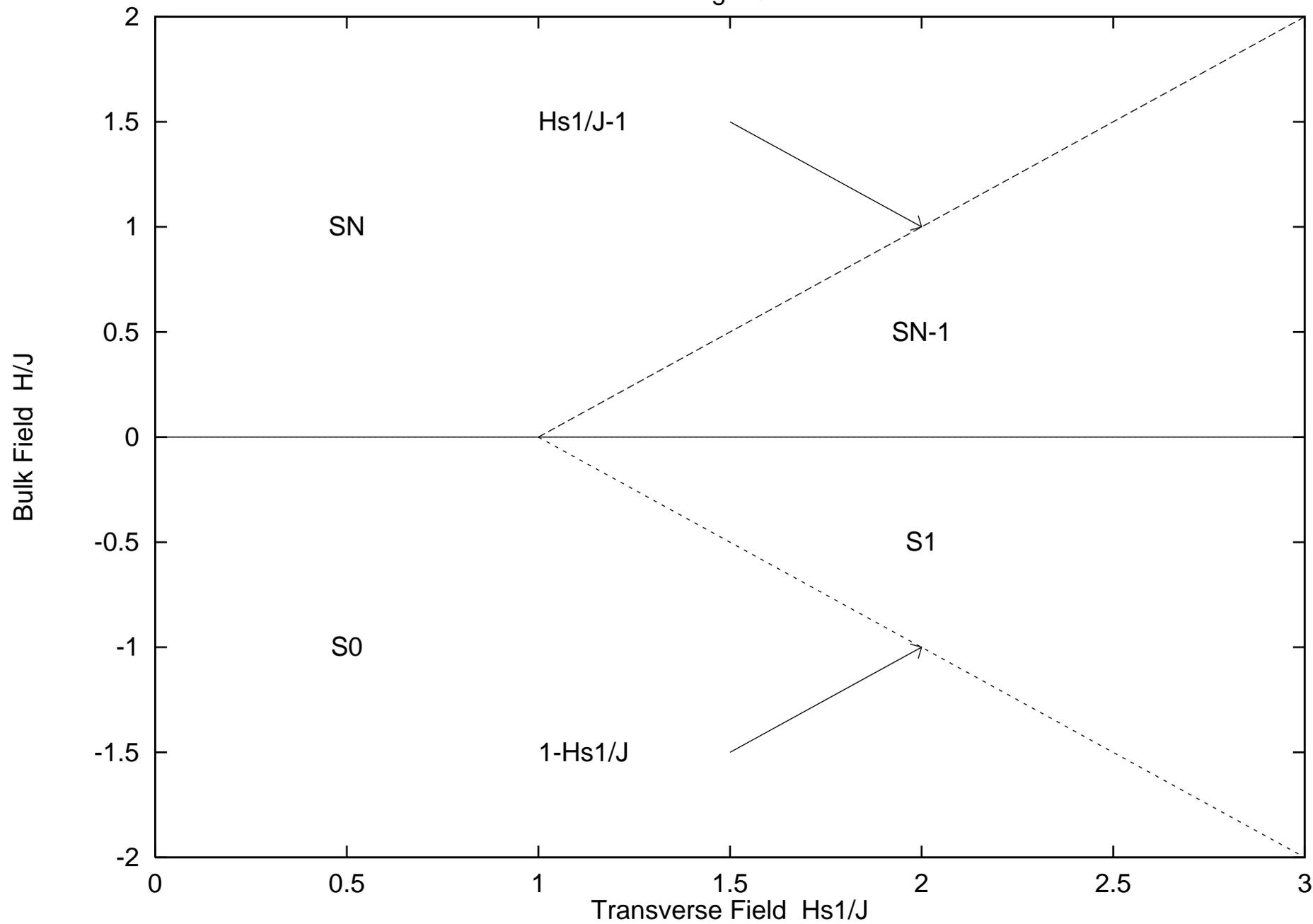
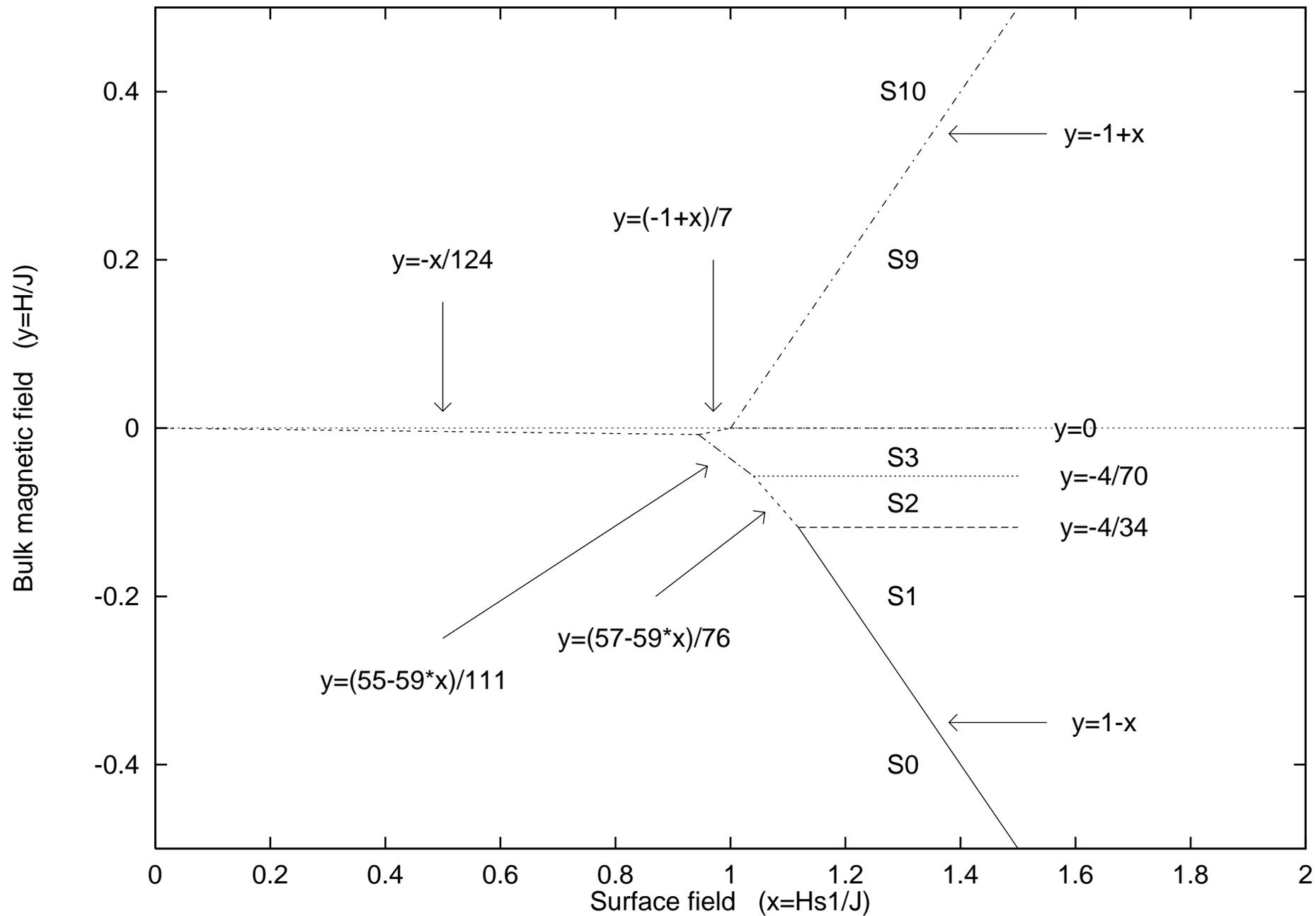
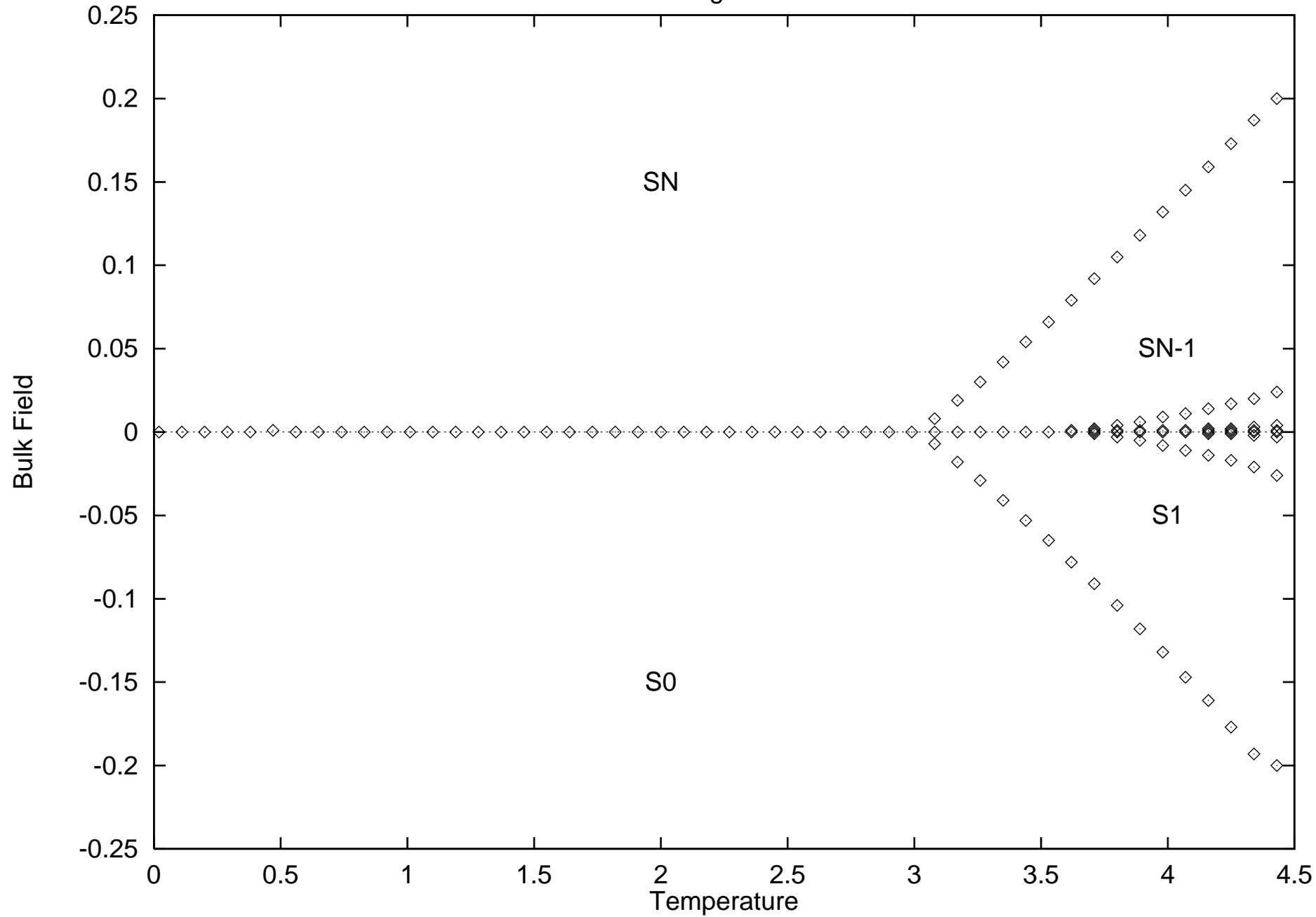
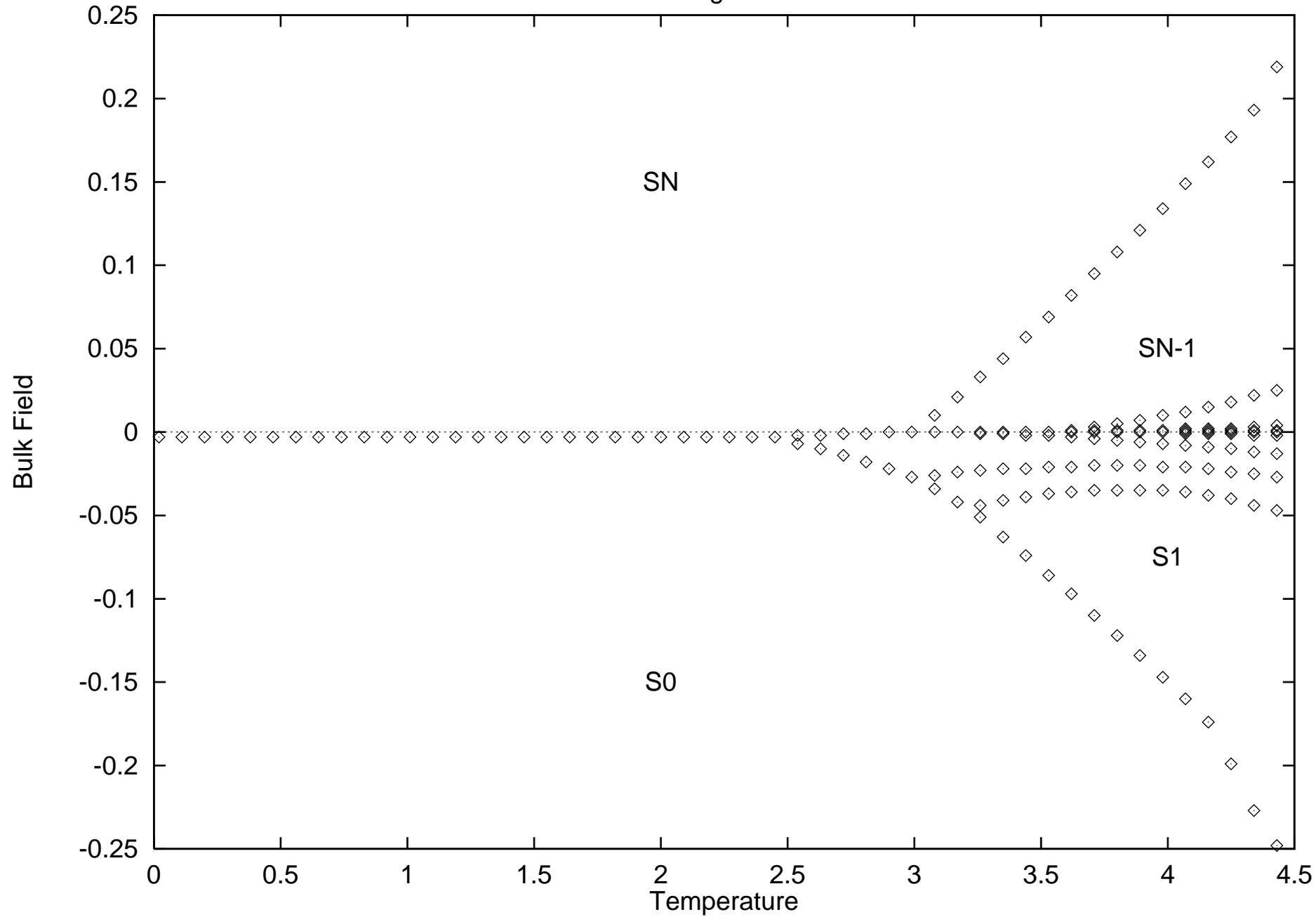
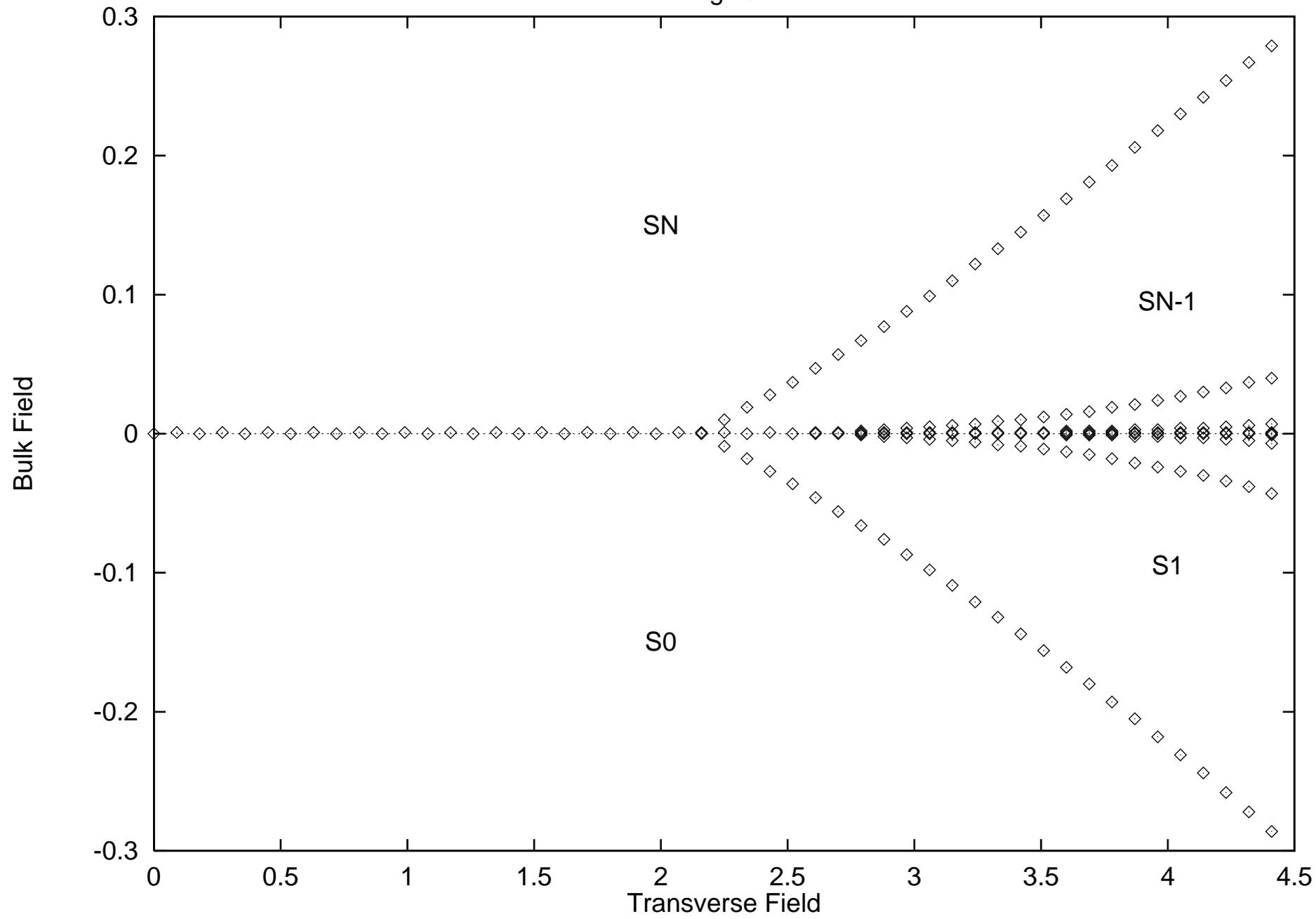
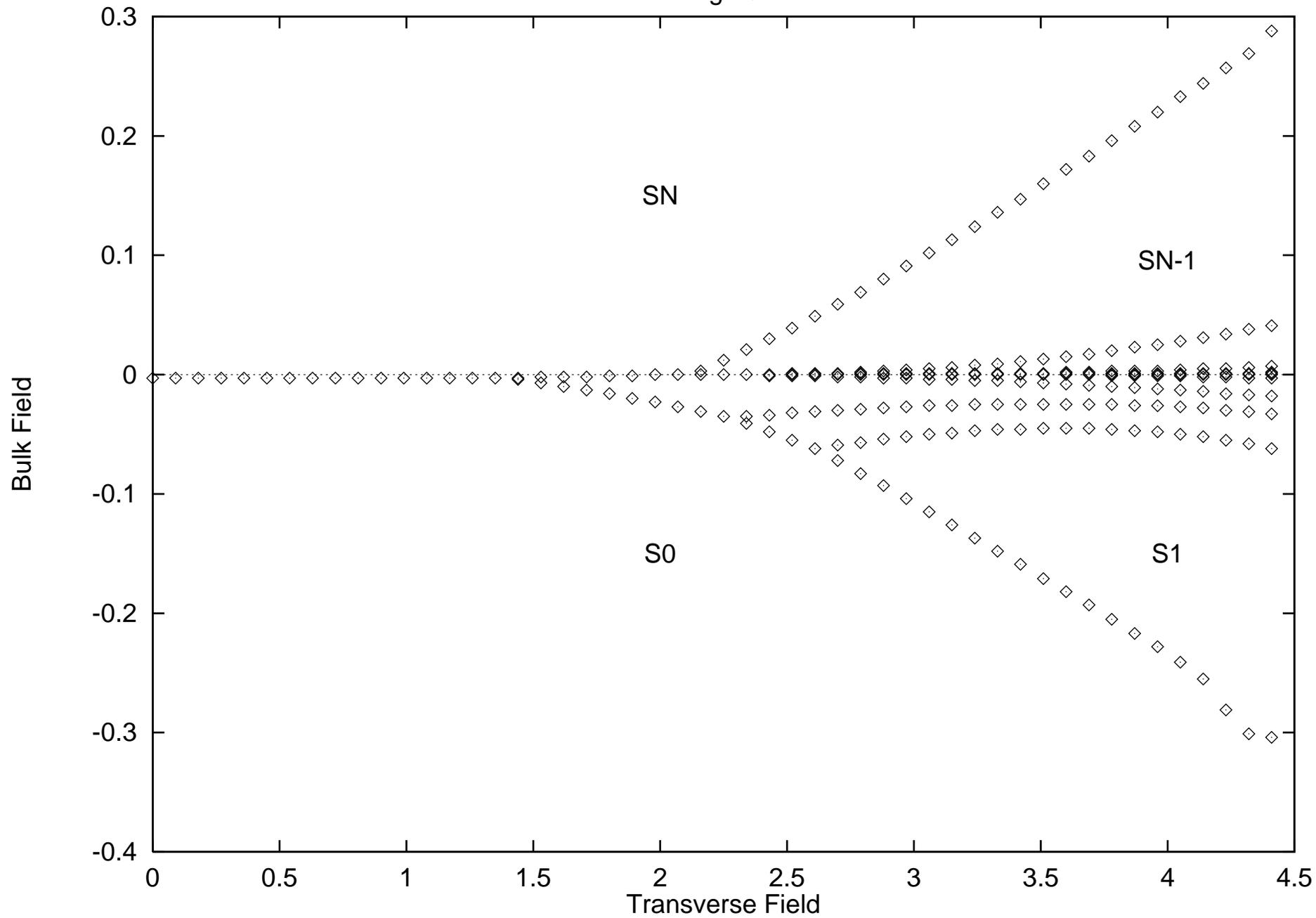
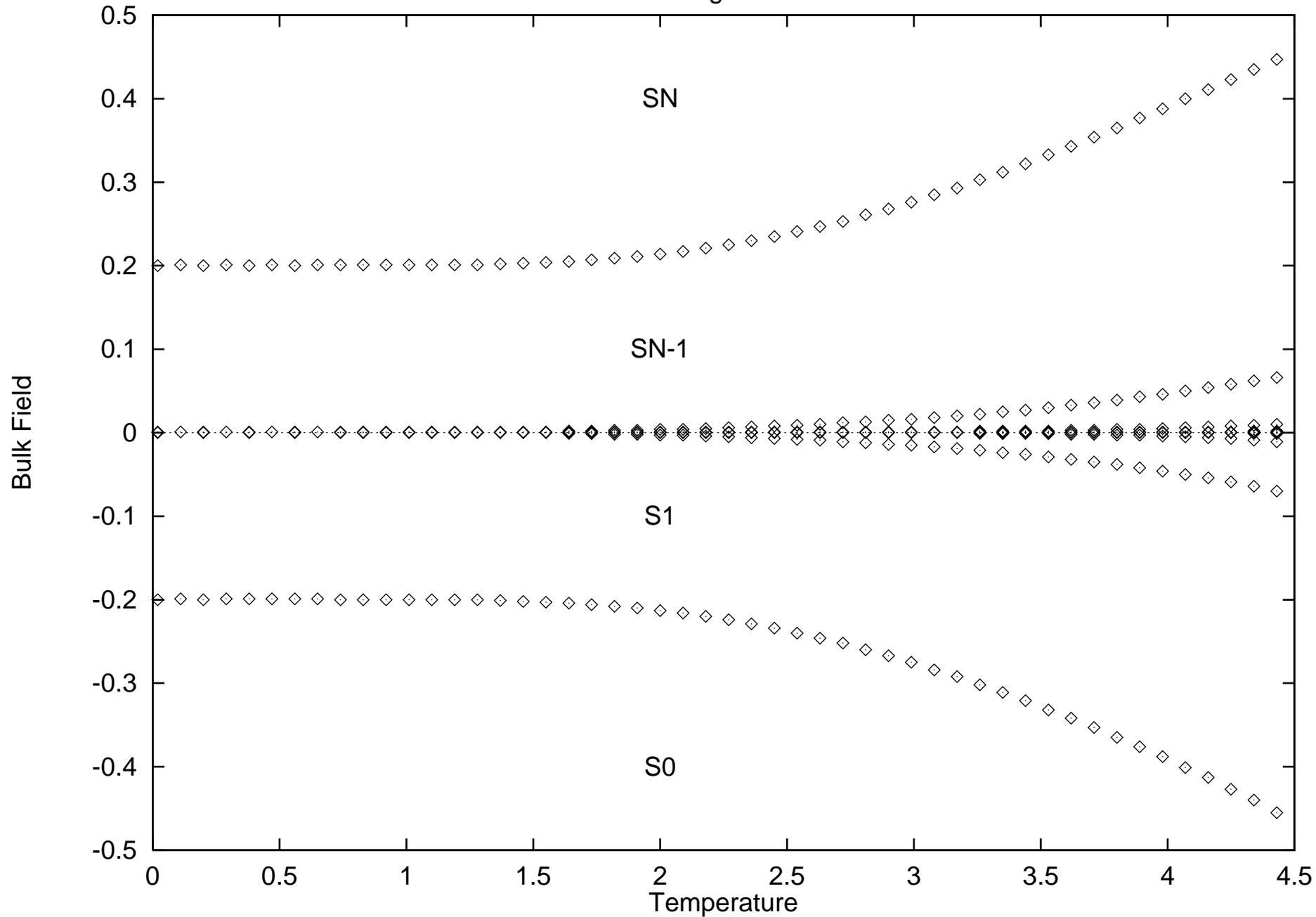
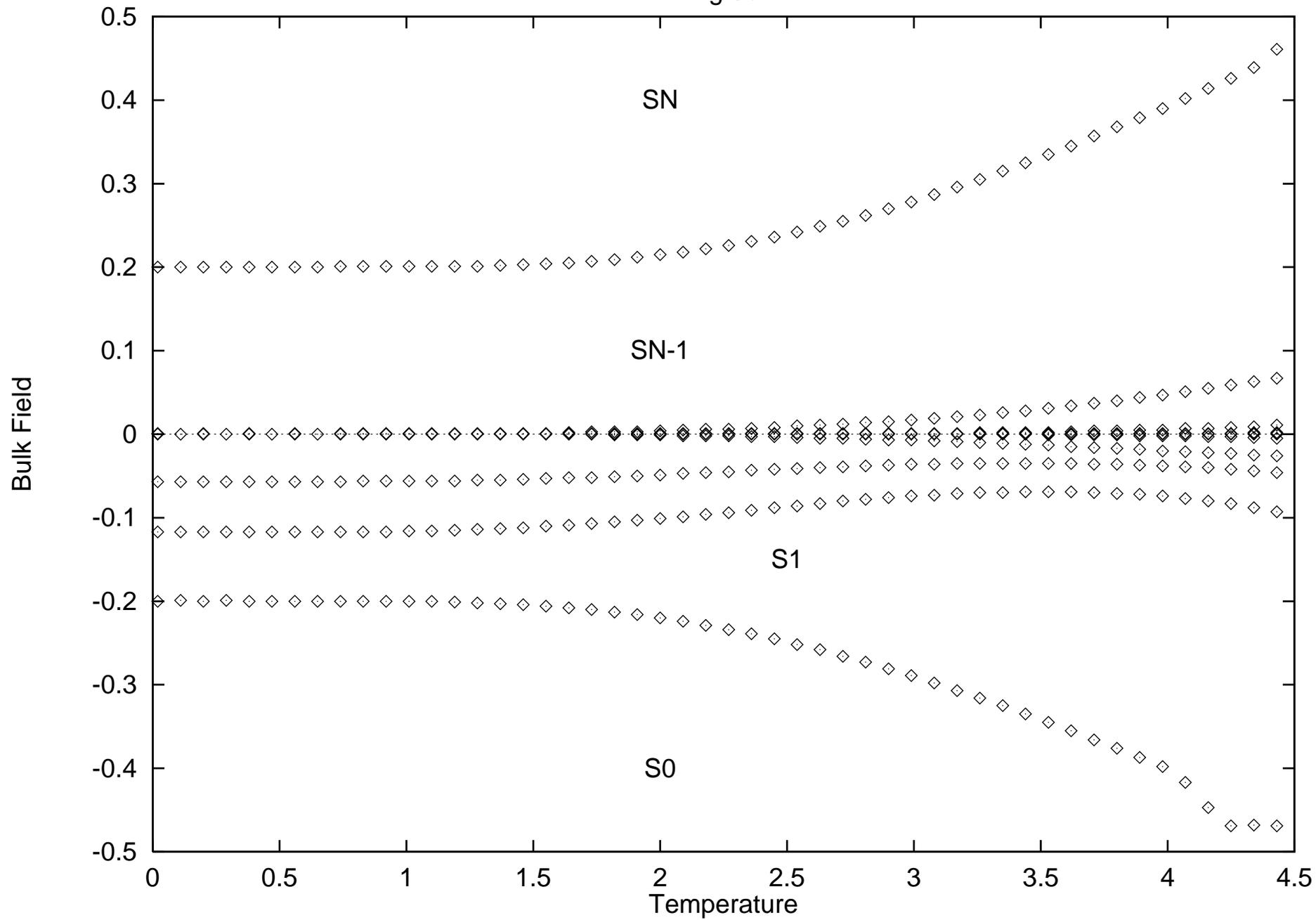
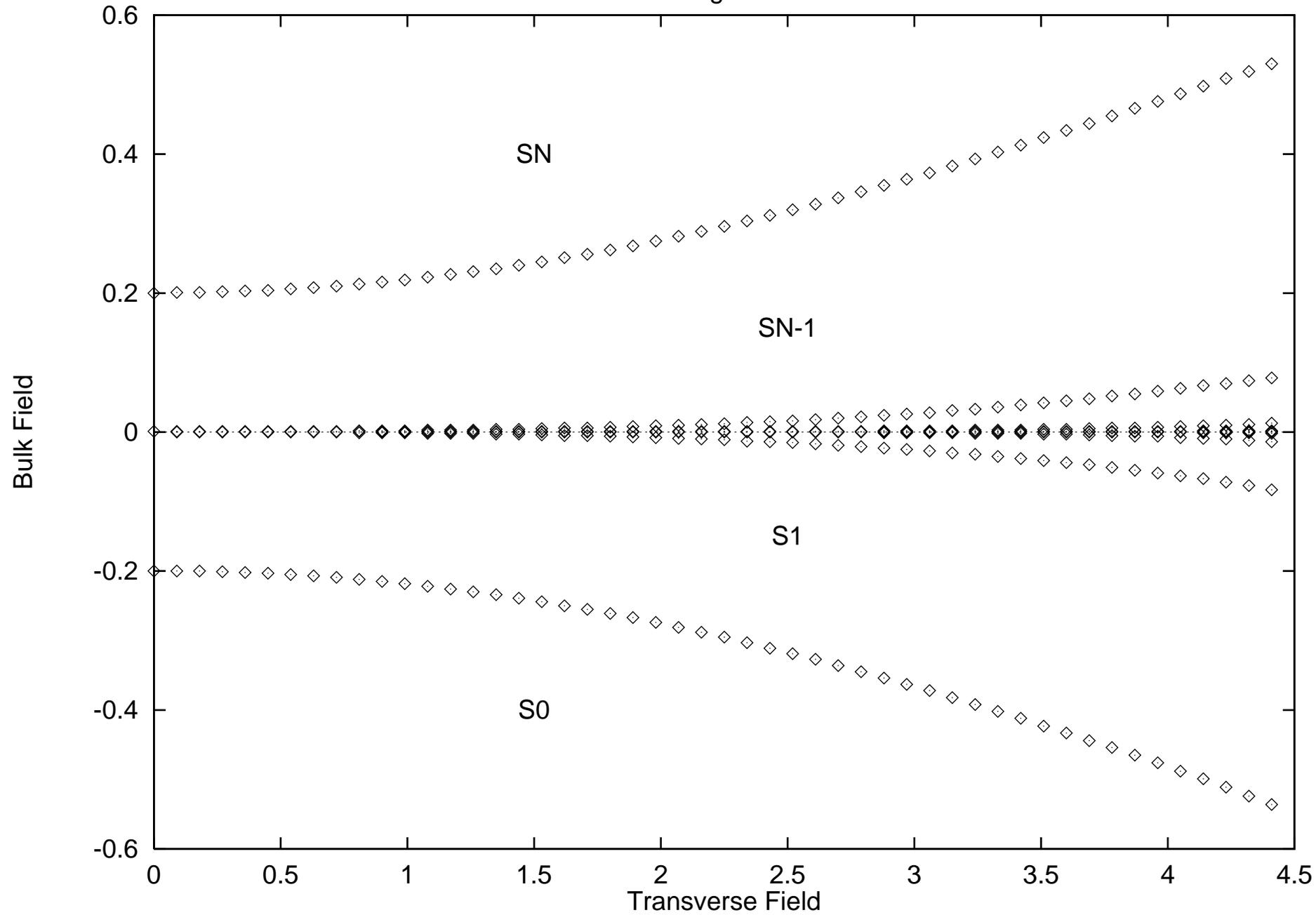
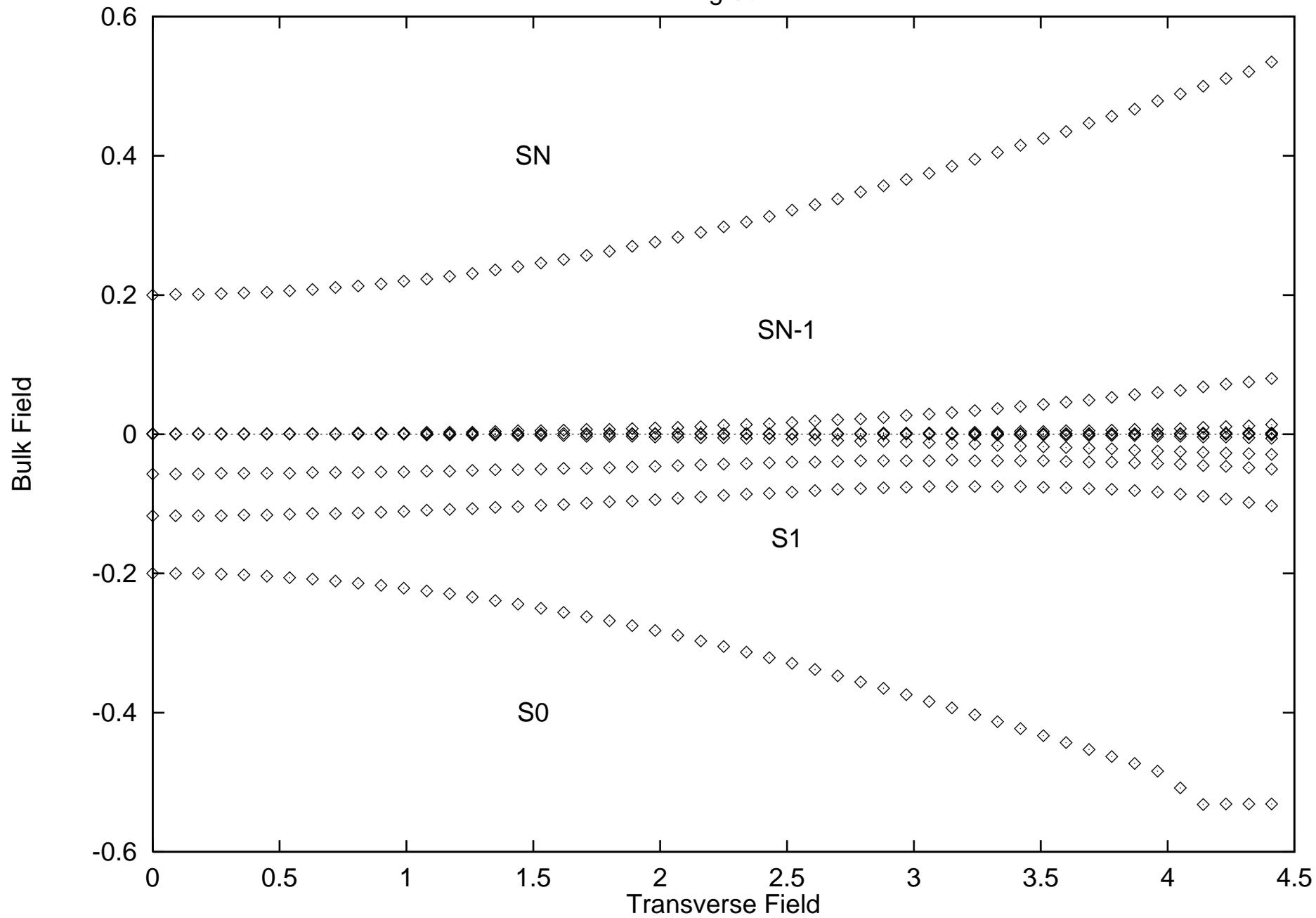
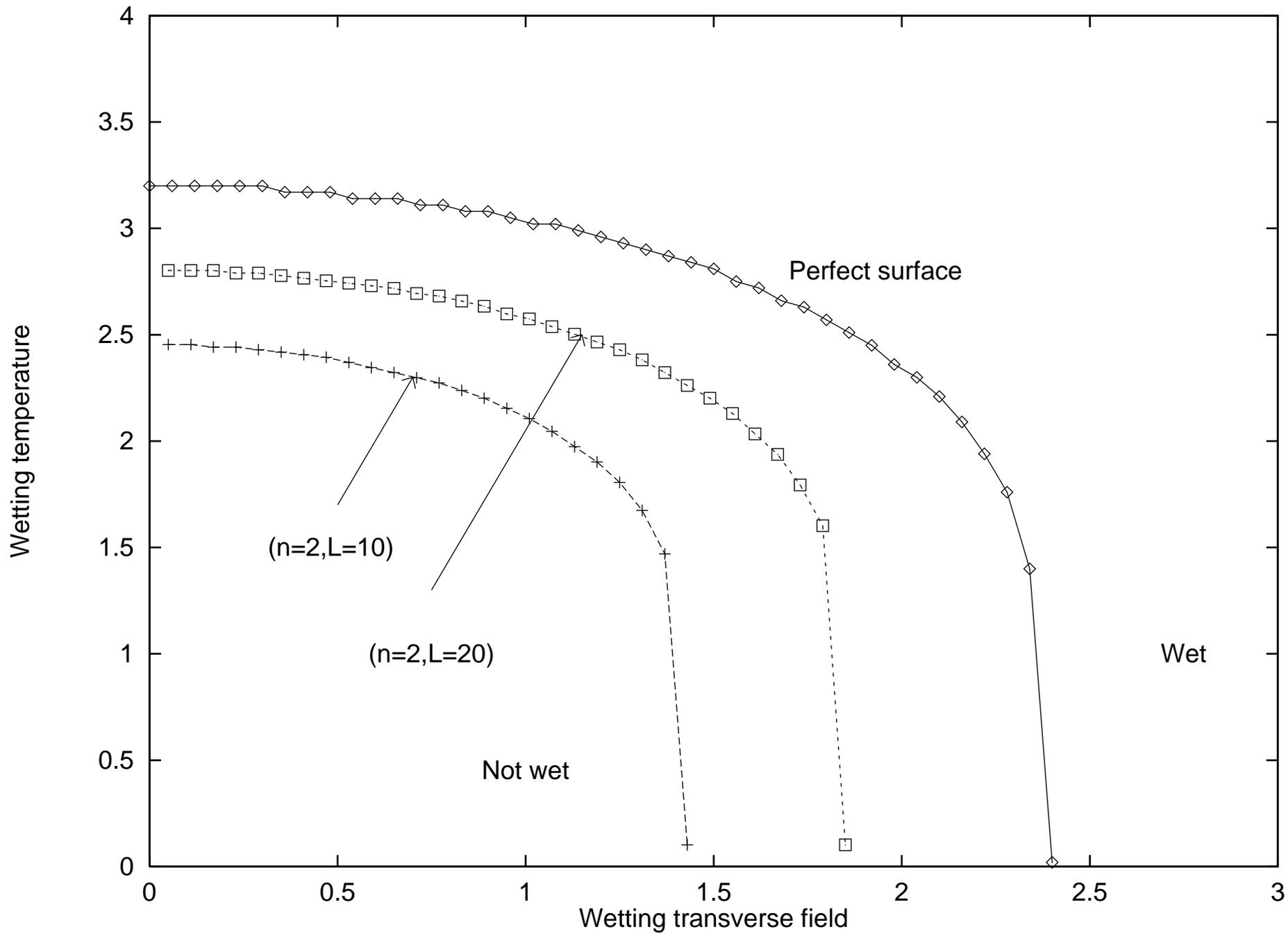
\end{document}